**Structural, magnetic, and nanoscale switching properties of BiFeO$_3$ thin films grown by pulsed electron deposition**

Hongxue Liu [a], Ryan Comes [a], Yonghang Pei [b], Jiwei Lu [a], Stuart Wolf [a,b]

[a] Department of Materials Science and Engineering, University of Virginia, Charlottesville, VA 22904
[b] Department of Physics, University of Virginia, Charlottesville, Virginia 22904

We report the epitaxial growth of BiFeO$_3$ by pulsed electron deposition and the resulting crystal quality, magnetic and nanoscale switching properties. X-ray diffraction shows high quality single phase, epitaxial (001) oriented films grown on SrTiO$_3$ (001) substrates. Both field and temperature dependent magnetic properties reveal an antiferromagnetic behavior of the films. For the film with a SrRuO$_3$ bottom electrode, an exchange-enhancement effect between antiferromagnetic BiFeO$_3$ and ferromagnetic SrRuO$_3$ was observed at low temperature. The piezoelectric force microscopy and switching spectroscopy measurements demonstrate the local domain switching process and suggest that the BiFeO$_3$ films are high quality ferroelectrics.

Multiferroics are interesting materials showing the simultaneous presence of multiple ferroic orders such as ferroelectricity and ferromagnetism.[1] The interplay of ferroic orders opens large potential applications in non-volatile memory and low power logic devices that can be controlled both electrically and magnetically.[2] BiFeO$_3$ (BFO) is one of the most promising candidate multiferroics with both magnetic and ferroelectric order well above room temperature required for practical device applications. It is ferroelectric up to the Curie temperature ($T_C$) of ~ 1100 K and is antiferromagnetic below its Néel temperature ($T_N$) of ~ 640 K.[3]

While BFO was first synthesized in the late 1950s,[4] the surging research interest around it and, in the larger context, multiferroics in general, took off after the report of large electric polarization and ferromagnetism observed in BFO thin films, along with the development and advancement of thin film growth techniques.[5] A large amount of work has been done to explore the growth of BFO and other multiferroic thin films with different deposition methods such as pulsed-laser deposition (PLD), sputtering, chemical vapor deposition, and molecular beam epitaxy (MBE), with PLD as the prevailing method.[5-8] Pulsed electron deposition (PED), a relatively new energetic condensation technique, offers a cost effective alternative to PLD. Unlike PLD, where the ablation process is critically dependent on the optical absorption coefficient of the target material, in PED, the ablation depends solely on the range of electrons in the target.[9] Thus PED also features a wider range of materials deposition capability, that can be appealing considering the increasing importance of material engineering through composite growth. Recently, Comes et al. demonstrated the high quality epitaxy of various oxide films using PED as the growth technique.[10] In this paper, we report the epitaxial growth of BFO by PED and discuss the resulting crystal quality, magnetic and nanoscale switching properties.

The BFO films were grown on (001) SrTiO$_3$ (STO) substrates with and without bottom SrRuO$_3$ (SRO) electrodes at a substrate temperature of 600 °C in a PED system (Neocera Inc.) that had a base pressure better than 5×10$^{-8}$ torr. The electron gun is energized with 12.5 keV and pulsed at a frequency of 5 Hz. All the growth was performed in pure oxygen with a chamber pressure of 18 mTorr using a polycrystalline BFO target with 15% Bi to accommodate possible Bi loss due to the high vapor pressure at elevated temperatures during the growth. Structural properties and film thicknesses were characterized using X-ray diffraction (XRD, Rigaku Smartlab with a single crystal germanium monochromator). During XRD scans each step was measured for at least 6 sec in order to achieve adequate statistics. The magnetic properties were characterized from 50 K to 360 K with a vibrating sample magnetometer (Quantum Design VersaLab). Magnetic fields were applied parallel to the film plane during susceptibility measurements. The atomic force microscopy (AFM), piezoresponse force microscopy (PFM) and local switching spectroscopy measurements were performed on a NT-MDT Solver and an Asylum Cypher scanning microscope.

Figure 1 shows the XRD patterns of the BFO films grown on STO (001) and SRO buffered STO (001). The BFO/SRO thicknesses are 13 nm/0 nm and 17 nm/27 nm, respectively. The distinct peaks of BFO, SRO and STO in the spectra suggest single-phase films and show no evidence of secondary phases within the detection limit of XRD. All the spectra feature only (001) type diffraction peaks, suggesting that the films are exclusively characterized by [001] epitaxial growth. Kiessig fringes are clearly observed, indicating a high quality epitaxy with smooth interfaces, uniform thickness and low defect density. The excellent film crystallinity is further confirmed by the full width at half maximum (FWHM) value of 0.07º of the rocking curve for the BFO (002) peak (insert of Fig. 1). For comparison, a range of 0.012º-0.26º of FWHM was reported for PLD grown BFO films on STO (001) and 0.35º-0.42º for sputtered films.[6,11]

The BFO (002) peaks clearly shift to lower angle positions compared to the 45.78° of bulk BFO,[3] signaling an out-of-plane (OP) tensile strain in the BFO films. This is induced by the in-plane (IP) compressive strain expected in the ultra-thin films due to the larger lattice constant of BFO compared to that of the STO substrate. Fig. 2(a) shows the XRD reciprocal space map (RSM) around the (103) reflection of the SRO buffered BFO film. The peak from the BFO is perfectly aligned on the horizontal axis with those of SRO and STO, indicating that both the SRO and BFO layers are fully strained and have the same IP parameters. The degree of IP orientation was accessed by the IP phi scan of the BFO films, as shown in Fig. 2(b). The phi peaks for the (101) reflection of the (001)-oriented domain occur at the same azimuthal phi angles as those for STO (101) reflection and are 90° apart from each other, which clearly indicates the presence of four fold symmetry along the [001] direction and epitaxy of BFO on (001) STO substrates.

In order to illustrate the magnetic ordering in BFO films, the temperature dependence and field dependence of the magnetization were measured. As shown in Fig 3(a), the temperature dependence of the magnetization of the BFO film directly grown on STO (001) first shows a slight decrease of magnetization with increasing temperature up to around 270 K and then shows an increase of magnetization as the temperature is further increased. The magnetization behavior below 270 K can be attributed to some paramagnetic defects from the film or the substrate, such as vacancy defects with unpaired electrons often observed in many oxide materials.[12,13] The magnetization above 270 K is characteristic of the temperature dependent magnetization of an antiferromagnetic material. The room temperature field dependent magnetization, as shown in the insert of Fig. 3(a), illustrates a linear field dependence, further confirming that the BFO film shows antiferromagnetic behavior. Due to the large diamagnetic background signal from the STO substrate, it is not possible to accurately determine the magnetic susceptibility of the BFO film. Nonetheless, These magnetic results clearly show the antiferromagnetic property of the BFO film. It should be pointed out that there were some previous reports of large magnetic moments at room temperature in BFO thin films and the weak ferromagnetism was attributed to the destruction of cycloidal spin order due to epitaxial strain.[5,6] Recent reports show that the parasitic phases such as $\gamma$-$Fe_2O_3$ contribute to the increased magnetic moments in thin films and the intrinsic magnetization of high quality BFO films is now thought to be near zero.[14-16] Our results agree well with the antiferromagnetism of BFO and is consistent with the XRD results suggesting single-phase BFO and the absence of any Fe-related clusters or parasitic phases in the BFO films.

The magnetic properties of the BFO film grown on the SRO buffered STO (001) substrate is shown in Fig. 3(b). The temperature dependent magnetization curve shows a ferromagnetic transition at about 150 K corresponding to the $T_C$ of the strained SRO layer,[17] and a similar increase of magnetization with increasing temperature above 250 K due to the antiferromagnetic behavior of BFO. While the room temperature field dependent magnetization shows a linear behavior, the magnetic hysteresis loop measured at 50 K is characteristic of the ferromagnetism in SRO and shows an enhanced coercivity after field cooling, compared to the coercivity of a SRO film of the same thickness grown under the same condition. This behavior of enhanced coercivity suggests an exchange-enhancement effect exists between antiferromagnetic BFO and ferromagnetic SRO at low temperature and is similar to that reported in BFO-$Fe_3O_4$ composite thin films.[18]

Fig. 4(a) - (c) show the AFM topography and OP and IP PFM images which clearly reveal the presence of irregularly shaped domain walls. The homogeneous OP phase image exhibits a monodomain pattern with one preferred polarization direction while the IP PFM phase image shows a clear two level contrast with domain polarization shifted by 180º. This is very similar to structures observed in ultra-thin BFO films grown on $La_{2/3}Sr_{1/3}MnO_3$ buffered (001) STO.[19,20] The domain morphology is quite different from the large striped domains reported in thick BFO films,[6] that may be attributed to the large epitaxial elastic strain energies in ultra-thin films and the large difference in film thickness as explained by the Landau-Lifshitz-Kittle scaling law.[19]

To demonstrate the polarization switch of the films, a dc bias was applied between the conducting AFM probe and SRO bottom electrode while scanning over the desired areas. Figure 5(a) shows OP PFM image and domain switching process of the BFO film under different poling bias. At -1 V and -2 V, the OP polarization still has the same direction as the spontaneous one and is thus not switched. At -3 V and -4 V, all ferroelectric domains are switched to an up-polarization state. The line profile of the piezoresponse further confirms this continuous domain switching process and suggests the good ferroelectric property of the BFO film.

The local ferroelectric switching properties were further studied using the PFM switching spectroscopy technique to measure the ferroelectric loops.[21] The applied sweep dc voltage is a triangular pattern starting from 0 to +6 V to -6 V to 0 and measurements are repeated 4 times to improve the signal to noise ratio. The averaged butterfly like piezo-amplitude curve and square piezo-phase hysteresis loop of the BFO film are shown in Fig. 5(b) and 5(c). The hysteretic behavior of the piezo-amplitude clearly shows two strain states at positive and negative voltages. The 180° change of the phase angle shown in the phase hysteresis loop verifies the complete domain polarization change, an indication of ferroelectric switching.

In summary, XRD shows single phase, epitaxial (001) oriented films grown on $SrTiO_3$ (001) substrates by PED. Together with the presence of clear Kiessig fringes and narrow FWHM, the results suggest the high quality and epitaxial nature of the films. The temperature dependent magnetic measurement and the linear field dependent magnetization curve show the antiferromagnetic order of the BFO films and indicate the absence of ferromagnetic parasitic impurities. The piezoelectric force microscopy and local switching spectroscopy measurements confirm the domain switching process and the good ferroelectric property of the BFO films.

**Acknowledgements**

This work was supported by DARPA under contract no. HR-0011-10-1-0072 and the Nano-electronics Research Initiative through NSF under contract no. DMR-08-19762. Ryan Comes also wishes to acknowledge support from the National Defense Science and Engineering Graduate Fellowship.

**Figure Captions**

Fig. 1. XRD patterns of BFO films grown on STO (001) and SRO buffered STO (001) at 600 °C. The BFO/SRO thicknesses are 13 nm/0 nm and 17 nm/27 nm, respectively. The insert shows the rocking curve for the (002) peak of the SRO buffered BFO film.

Fig. 2. (a) Reciprocal space mappings around the (103) reflections, and phi scan of the (b) BFO and (c) STO (101) reflections of the SRO buffered BFO film.

Fig. 3. Temperature dependent and field dependent (insert) magnetization of BFO films grown on (a) STO (001) and (b) SRO buffered STO (001). The diamagnetic signals from the STO substrates are not subtracted except in insert (b).

Fig. 4. (a) AFM topography, (b) OP and (c) IP PFM images of the SRO buffered BFO film.

Fig. 5. (a) Domain switching process after polling at different bias voltages, and local switching spectroscopy PFM (b) amplitude and (c) phase curves of the SRO buffered BFO film.

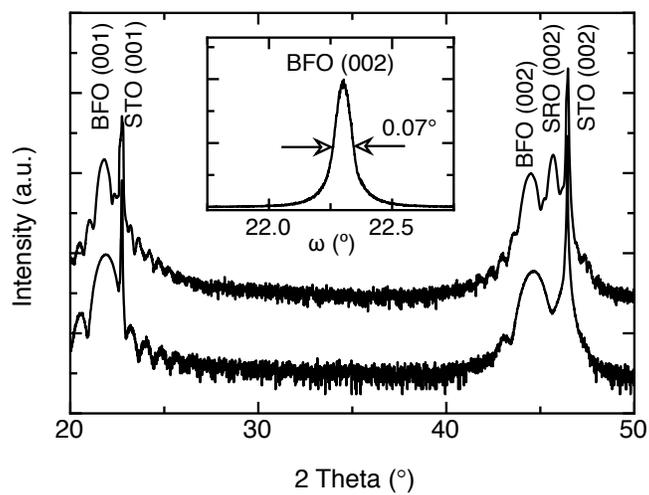

Fig. 1. Liu, et al.

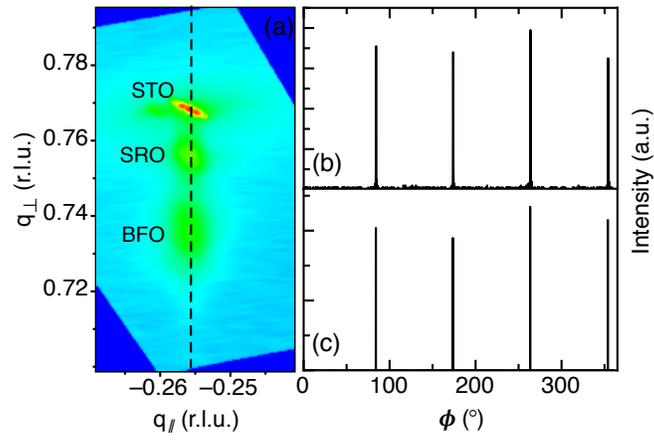

Fig. 2. Liu, et al.

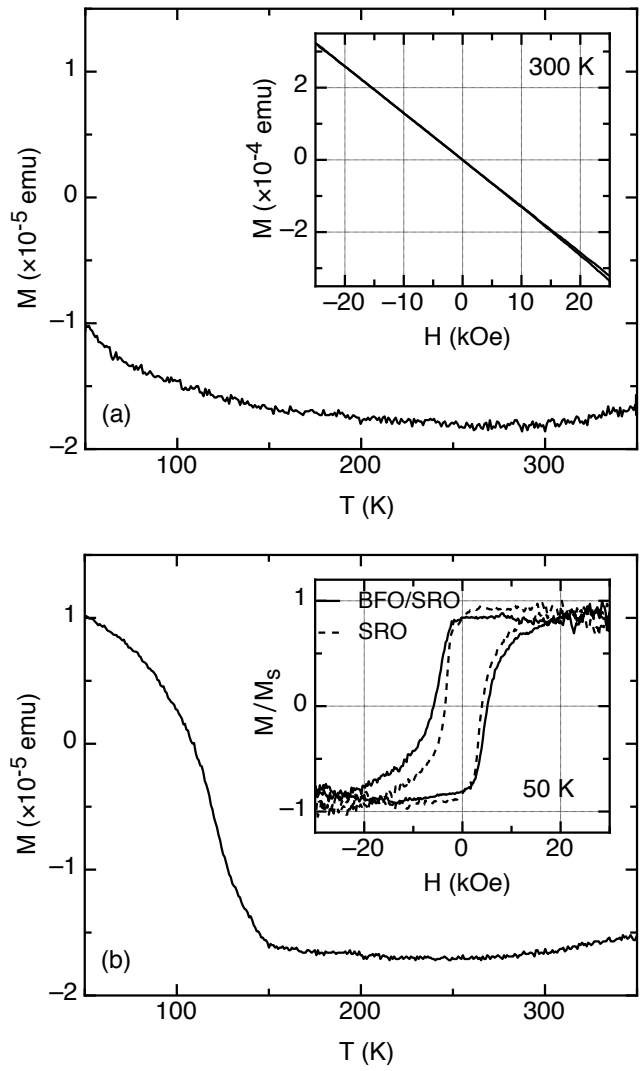

Fig. 3. Liu, et al.

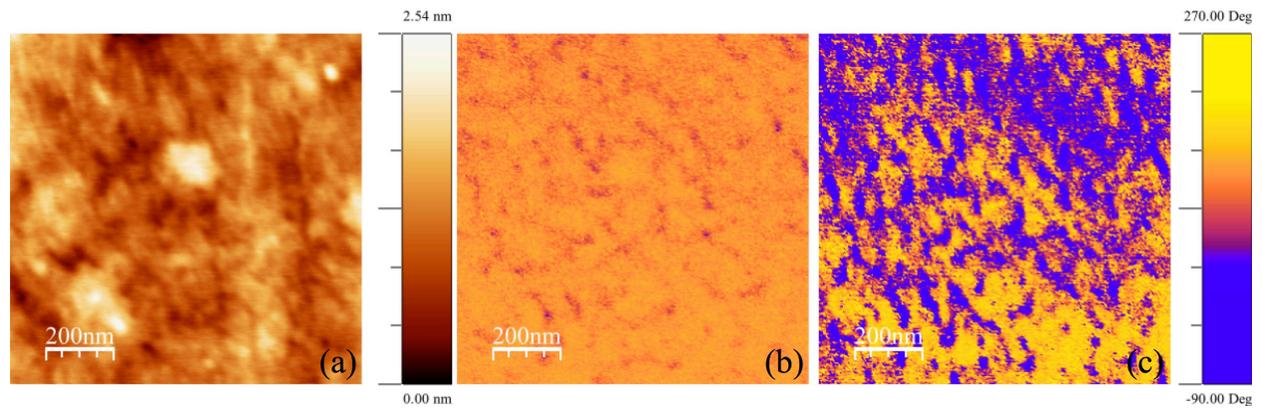

Fig. 4. Liu, et al.

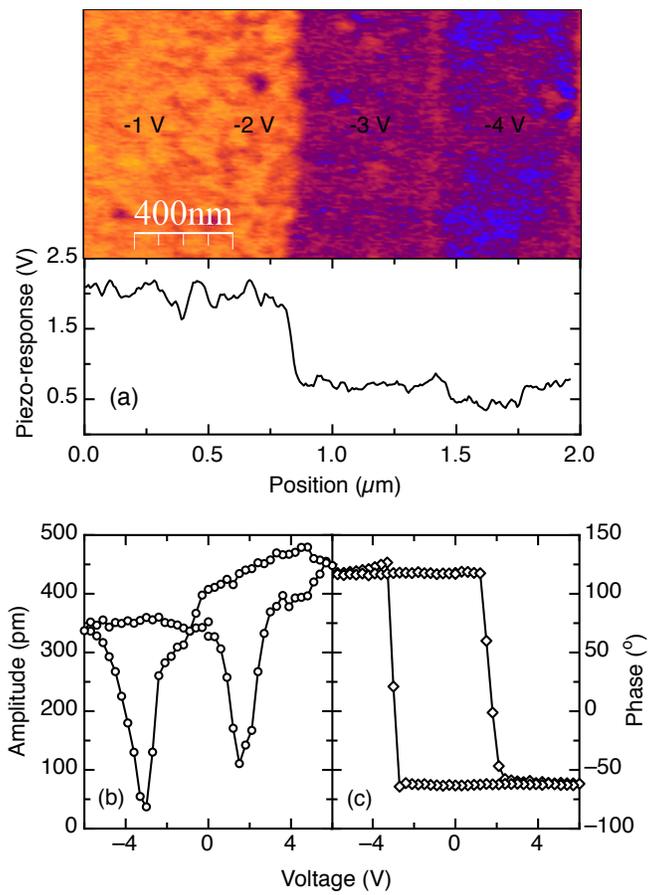

Fig. 5. Liu, et al.